\documentclass[debug]{epl-preprint}
\usepackage{graphics}

\begin{document}
\title{Zero Temperature Insulator-Metal Transition in Doped Manganites}
\shorttitle{ $T=$0 MIT in Doped Manganites} \shortauthor{G. V. Pai
{\it et al.}}
\author{ G. Venketeswara Pai\inst{1,2}\thanks{Email : venkat@ictp.trieste.it},
 S. R. Hassan\inst{1, 3}\thanks{Email:hassan@physics.iisc.ernet.in}
 H. R. Krishnamurthy\inst{1,3}\thanks{Email:hrkrish@physics.iisc.ernet.in},
 and T. V. Ramakrishnan \inst{1,3}\thanks{Email: tvrama@physics.iisc.ernet.in }}
\institute{ \inst{1} Centre for Condensed Matter Theory,
Department of Physics, Indian Institute of Science, Bangalore - 560012, India\\
\inst{2} The Abdus Salam International Centre for Theoretical
Physics, Strada Costiera 11, 34014 Trieste, Italy\\
\inst{3} Jawaharlal Nehru Centre for Advanced Scientific
Research, Bangalore-560064, India}
 {Colossal Magnetoresistance}

\maketitle
\begin{abstract}
We study the transition at $T=0$ from a ferromagnetic insulating
to a ferromagnetic metallic phase in manganites as a function of
hole doping using an effective low-energy model Hamiltonian
proposed by us recently. The model incorporates the quantum nature
of the dynamic Jahn-Teller(JT) phonons strongly coupled to
orbitally degenerate electrons as well as strong Coulomb
correlation effects and leads naturally to the coexistence of
localized (JT polaronic) and band-like electronic states. We study
the insulator-metal transition as a function of doping as well as
of the correlation strength $U$ and JT gain in energy $E_{JT}$,
and find, for realistic values of parameters, a ground state phase
diagram in agreement with experiments. We also discuss how several
other features of manganites as well as differences in behaviour
among manganites can be understood in terms of our model.
\end{abstract}

The colossal magneto-resistance(CMR) exhibited by manganites
($Re_{1-x}A_x MnO_3\,, Re = La, Nd, Pr $ {\it etc.} and $A = Sr,
Ca, Ba$ {\it etc.}) for a range of hole doping $x$ around $x \sim
0.3$ and near the Curie temperature $T_c$, where they undergo a
transition from a low temperature ferromagnetic metallic phase to
a high temperature paramagnetic insulating phase, has led to a
great deal of  interest\cite{reviews,dagotto} in these systems,
which also show a variety of other interesting phenomena such as
charge and orbital ordering and incipient phase separation. The
interplay of orbital degeneracy of the itinerant $e_g$ electrons
of $Mn$, their coupling to lattice degrees of freedom, especially
to degeneracy removing Jahn-Teller(JT) phonons, strong Coulomb
correlation effects and related Hund's rule coupling (between the
$e_g$ electrons and the $t_{2g}$ core spins of $Mn$) are believed
to be responsible for these phenomena, but achieving a detailed
theoretical understanding has been a major challenge.

We have recently proposed\cite{tvr-nat,theses} a new effective
low-energy Hamiltonian starting from the  two qualitatively
different coexisting vibron\cite{vibron} states at the each site
of the lattice, one consisting of localized JT
polarons\cite{polaron_expt}, which we label $\ell$, and the other,
which we label $b$, dispersive and forming a broad band as
outlined below.  In the presence of strong correlation $U$ and
ferromagnetic Hund's rule coupling $J_H$ on site between $e_g$ and
$t_{2g}$ spins this leads \cite{tvr-nat,theses} to a consistent
description of many known features of these systems such as the
finite temperature insulator metal transition (IMT) near $T_c$ for
a range of $x$, CMR, the existence of a ferromagnetic insulating
phase at low doping, good metallic behaviour of electron doped
systems, etc., and a heuristic understanding of several other
features, such as the isotope effect on $T_c$ and the {\it
two-phase coexistence} seen by a variety of experimental probes
over a range of $x$ and $T$ \cite{dagotto,2-phase}. In this paper,
we present in detail our theory for the {\em {ground state
behaviour of manganites in the ferromagnetic state}} as a function
of doping $x$ (for  $x < x_{co}$ where charge ordering occurs) in
terms of the Falicov-Kimball model (FKM) \cite{fkm} involving the
$\ell$ and $b$ states.

The un-doped compound, eg. $La Mn O_3$ is an $Mn-O$ bond (JT)
distorted but structurally ordered,  A-type anti-ferromagnetic
insulator with ferromagnetic order of the $t_{2g}$ spins in plane
(defined by the JT distortion pattern) and anti-ferromagnetic
order perpendicular to it.  On hole doping, say with $Sr$,
anti-ferromagnetism disappears at $x=x_{c1}=.08$ and the ground
state is a ferromagnetic insulator till $x=x_c=.16$, beyond which
it is a ferromagnetic metal. For $LaCa$, $x_{c1}=0.1$ and
$x_{c}=0.18$ whereas for $NdSr$,  $x_{c1}=0.18$ and $x_{c}$ is not
accurately known. The occurrence of the fully ferromagnetic
insulating  phases and the IMT at such large values of $x$ cannot
be understood in a model for manganites with only $e_g$ electron
double exchange caused by $J_H$ \cite{reviews,d-e}. For, since
doping generates holes (unoccupied sites) in the $e_g$ band, one
expects a metal especially when the $t_{2g}$ spin alignment is
ferromagnetic, since the $e_g$ electrons can then move without
hindrance. This is true even for  large $U$.

We outline below (see ref.\cite{tvr-nat} for details) how the FKM
\cite{fkm} describing the correlated $\ell$ and $b$ states arises
in the context of manganites at $T=0$ from a conventional lattice
model with two $e_g$ orbitals per site, large electron JT phonon
coupling $g$ and perfect spin alignment resulting from a {\em{new
ferromagnetic 'virtual double exchange' coupling}} $J_F$. We treat
it using dynamical mean field theory (DMFT) \cite{dmft}, and show
that the $b$ band has a {\em{reduced effective width}}, $2D$,
roughly equal to $\sqrt{x}$ times its bare width $2D_0$ for large
$U$. Hence, below a critical doping $x_c$ the $b$ band bottom lies
above the localized JT polaronic $\ell$ levels of energy
$-E_{JT}$, which are then the only ones occupied, leading to an
insulator. For larger $x$, the $b$-band bottom lies below the
$\ell$ level, so the system is metallic. The ferromagnetism is
largely due to $J_F$ (with a small contribution from conventional
double exchange in the metallic phase). Thus the observed $T=0$
pattern of phases and phase transitions follows naturally from our
picture \cite{comment1}. We discuss the ground state properties of
the FKM as a function of the hole doping ($x$) and the model
parameters, namely $U$, $D_0$ and $E_{JT}$, and show that for
realistic values of these parameters, the calculated phase
boundary is in good agreement with experimental trends. We
conclude by discussing some other implications of our work in
regard to experiments. A detailed discussion of the finite
temperature properties of our model, including the ferro metal-
para insulator transition and the CMR, is presented
elsewhere\cite{tvr-nat}.

We start with a model with two degenerate $e_g$ orbitals per site
and strong degeneracy breaking electron JT phonon coupling $g$.
For large $g$, at each site there will be one {\it vibron}
solution \cite{vibron}, labelled $\ell$, with its energy reduced
for single occupancy due to a large JT distortion by an amount
$E_{JT}= (g^2/2K) \simeq 1 eV$ \cite{MillisRoyalSc} where $K$ is
the force constant of the $JT$ phonon mode. The orthogonal vibron
solution \cite{vibron}, labelled $b$ , is not JT distorted, and
hence has no gain in JT energy.  Inter-site hybridization of the
$\ell$ states is reduced by the phonon overlap  or Huang-Rhys
factor\cite{HuangRhys} $\eta = exp (-E_{JT}/2\hbar \omega_0)$,
which for manganites is $\sim (1/200)$ since
$(E_{JT}/\hbar\omega_0)\simeq 10$. As a first approximation, we
neglect this altogether and treat the $\ell$ states as site
localized. Inter-site hybridization amongst the $b$ electrons is
not suppressed. They hence form a broad band. In the presence of
large $U$ the $b$ states have their largest amplitudes on hole
sites ($x$) as they are strongly repelled from the polaronic sites
$\ell$. In the presence of large $J_H$ the spins of both $\ell$
and $b$ electrons are aligned parallel to the $t_{2g}$ core spins.

An additional consequence\cite{tvr-nat} of the existence of JT
distorted, localized $\ell$ states is that {\em{virtual,
adiabatic}} hopping processes involving them {\em{in the presence
of large $U$ and $J_H$}} (where an $\ell$ electron at site $i$
quickly hops to an {\it empty neighbour} $j$ and back, with an
intermediate state energy cost of $2E_{JT}$ due to the unrelaxed
lattice distortion at $i$) give rise to a {\em{new, major, doping
dependent ferromagnetic nearest neighbour exchange coupling}}
$J_F$ between the $t_{2g}$ core spins of order $t^2 (1-x)
x/(2E_{JT} S^2)$. In the  ground state for $x > x_{c1}$  when this
interaction dominates the anti-ferromagnetic super-exchange, the
$t_{2g}$ are fully ferromagnetically polarized; then so also are
the $e_g$ (both $\ell$ and $b$) spins due to the large $J_H
(>>t)$. Hence the spin degrees of freedom are frozen.

Thus we are led to effectively spin-less localized $\ell$ and
mobile $b$ electrons with a strong local Coulomb repulsion $U$.
The relevant effective Hamiltonian is just the Falicov-Kimball
model (FKM) \cite{tvr-nat,theses,fkm} , given by
\begin{equation}
H_{eff} = - \sum_{\left< ij \right>} \bar{t}_{ij} b^{\dag}_i b_j
-(E_{JT} + \mu) \sum_i \ell^{\dag}_i \ell_i - \mu \sum_i
b^{\dag}_i b_i + U \sum_i n_{bi} n_{{\ell}i} \label{eq-heff}
\end{equation}
Here $ b^{\dag}_i$ and $\ell^{\dag}_i$ create the band and
localized polaronic states described above at site $i$, and
$\bar{t}_{ij}$ are effective, orbitally averaged\cite{co-phases}
inter-site hopping amplitudes for the $b$ states. The chemical
potential {\em{$\mu$ is determined by the doping-dependent
constraint}} that the total number of $e_g$ electrons is $(1-x)$
per site. The relevant parameter regime of $H_{eff}$ we are
interested in for manganites corresponds to $\bar{t}\sim 0.2 \,
eV$ \cite{Satpthy}, $U$  very large ($\sim 5 - 8 \, eV$
\cite{Satpthy}) and $E_{JT}$ in the range $0.5 - 1.0 \, eV $
\cite{Satpthy,MillisRoyalSc}.

While there are no known techniques for exactly solving $H_{eff}$,
a dynamical mean field theory(DMFT) treatment of it can be carried
out exactly \cite{dmft}. In this approximation, which is exact in
infinite dimensions, and quite accurate for three
dimensions\cite{dmft}, the lattice problem is mapped to a single
site problem embedded in a self consistent effective medium or
electron bath that represents all the other sites of the lattice.
We assume that the system is homogeneous\cite{co-phases}. The $b$
electron self energy $\Sigma_{ij}(\omega)$ due to the interactions
$U$ is site local i.e. $\Sigma_{ij}(\omega) = \delta_{ij}
\Sigma(\omega)$, and is determined from the single site or local
effective action
\begin{eqnarray}
{\cal{S}}_{eff} &=& -\int^\beta_0 \int^\beta_0 d \tau d \tau'
b^{\dag}(\tau) {\cal{G}}^{-1}(\tau - \tau') b(\tau') + \beta
(-E_{JT} -\mu)n_{\ell} + U n_{\ell}\int^\beta_0 d \tau n_b(\tau)
\label{eq-seff}
\end{eqnarray}
Here $b^{\dag}(\tau)$ and $b(\tau')$ are fluctuating fermionic
Grassmann fields, $n_{\ell}=$ 1 or 0 corresponding to the $\ell$
state being occupied or not, and $\cal{G}(\tau)$ is the bare on
site local propagator in the presence of the effective medium. The
local partition function, obtained by summing  $exp (- S_{eff})$
over all the configurations, is expressible as $Z_{local} = Z_0 +
Z_1$. Here $Z_0, Z_1$ are {\it constrained partition functions}
corresponding to $n_\ell = 0, 1$ respectively, and are calculable
using standard techniques\cite{dmft} in terms of ${\cal
G}(\omega^+)$,the Fourier coefficients of ${\cal G}(\tau)$
analytically continued to real frequencies. Explicitly, $Z_0
\equiv exp (\beta \alpha_0 (T))\, ; Z_1= exp (\beta \alpha_1 (T))
\times exp \left[- \beta (-E_{JT} -\mu)\right]$; where
\begin{equation}
\alpha_m (T) =\pi^{-1} \int d\omega n^{-}_{F}(\omega-\mu) \, Im
\left\{\ell n ({\cal G}^{-1}({\omega}^+) - U
\delta_{m1})\right\}\,,
\end{equation}
$n^{-}_{F}(\omega)$ being the Fermi function $\left[ 1+ exp (\beta
\omega) \right]^{-1}$.  The local single particle Green's function
$G(\omega^+)$ is given by
\begin{equation}
G(\omega^+) = -\langle bb^+\rangle_{S_{eff}} = w_0 {\cal
G}(\omega^+) + w_1 ({\cal G}^{-1} (\omega^+)-U)^{-1} \label{eq-G}
\end{equation}
with $w_1 \equiv Z_1/(Z_0 + Z_1)= \bar{n}_\ell$ and $w_0 =
(1-w_1)$ being the annealed probabilities for the $\ell$ state
being occupied and empty respectively. The condition that the
average $e_g$ occupancy ${\bar n}_{\ell }+{\bar n}_{b}=(1-x)$
determines the chemical potential $\mu$. A re-normalized or
effective $\ell$ electron energy can be defined by writing $w_1 =
n^{-}_{F}(\epsilon^*_\ell -\mu)$, whence $\epsilon^*_\ell  =
-E_{JT} +\alpha_0 (T) -\alpha_1 (T)$. Clearly, at $T=0$, $\mu$ is
necessarily pinned to $\epsilon^*_\ell $ as long as $\bar{n}_\ell$
is non zero.

In the DMFT \cite{dmft} Eq. \ref{eq-G} relating $G(\omega^+)$ and
${\cal G}(\omega^+)$ for our model has to be supplemented  by two
other equations: namely the Dyson equation, $G^{-1} (\omega^+) =
{\cal G}^{-1} (\omega^+ ) -\Sigma (\omega^+)$, and the self
consistency relation which expresses the local Green's function in
terms of the lattice Green's function, as $G(\omega^+) = \int
d\epsilon D_0 (\epsilon) / (\omega^+ + \mu - \epsilon- \Sigma
(\omega^+) )$, where $D_0(\epsilon)$  is bare density of states
(DOS) for the b-band. These coupled equations for ${\cal G}, G$
and $\Sigma$ have to be solved self-consistently, and typically
numerically, to obtain all the quantities of direct physical
interest.  The self consistency relation becomes algebraic,
considerably simplifying such calculations, for a semicircular
DOS, i.e., for $D_0(\epsilon)=(2/\pi D_{0}^2)\,\sqrt{D^2_0
-\epsilon^2}$ where $D_0$ is the half bandwidth. This DOS, exact
for the Bethe lattice in infinite dimensions \cite{dmft}, is
fairly accurate for our model in 3d, especially for trends and
magnitude estimates \cite{real-band}. Hence we confine ourselves
to the semicircular DOS results in this paper. In this case, $G(z)
= 2/(z+\sqrt{z^2 -D_{0}^2})$ where $z\equiv(\omega^+ + \mu
-\Sigma(\omega^+))$ is complex.  Using this result, Eq.
\ref{eq-G}, the Dyson equation, and the equations for $w_0,w_1$ we
have numerically solved the DMFT equations self-consistently for a
wide range of values for the correlation $U$, JT polaronic energy
$E_{JT}$, bare bandwidth $2D_0$ and doping $x$. The results are
discussed below. A broad perspective of the trends in these can be
obtained in the simpler limit when $U\rightarrow \infty$. Hence we
discuss these first.

In the $U\rightarrow\infty$ limit, the DMFT equations stated above
(for the semicircular DOS) can be solved analytically. The local
Green's functions have the simple form $$G(\omega^+)=2w_0 /\left(
\omega^+ + \mu + \sqrt{(\omega^+ + \mu)^2- D^2} \right) =
w_{0}{\cal G} (\omega^+)$$ with  $D \equiv \sqrt{w_0}D_0$. The
local spectral function or re-normalized DOS of the $b$-band is
simply $\rho(\omega)= 2\sqrt{D^2 - (\omega + \mu) ^2}/(\pi
{D_0}^2)$, i.e., once again of the semi-circular form, but with a
reduced effective bandwidth $2D$, and reduced weight $w_0$. At
$T=0$,\, $\bar{n}_b$, $w_0 = 1 - \bar{n}_\ell $ and
$\epsilon_\ell^*$ can hence be evaluated from (numerically
solving) the equations:
$$\epsilon^{*}_{\ell} =-E_{JT}+(\mu {\bar n}_b /w_0) ; $$
$${\bar n}_b = (w_0-x) = (w_0/\pi)\left\{ sin^{-1}(\mu/D) +(\pi/2)
+ (\mu / D) \sqrt{1-(\mu /D)^2} \right\} \theta(\mu+D) .$$ In
addition, when ${\bar n}_{\ell}\ne 0$, there is the pinning
condition $\epsilon^{*}_\ell = \mu $. These equations have the
self consistent solution $\bar{n}_b=0,\,\, \epsilon_\ell^*
=-E_{JT} = \mu$, for $ E_{JT} > D \equiv \sqrt{w_0} D_0 =
\sqrt{x}D_0 $, since $w_0 = x$ for $\bar{n}_{b} =0$.  Thus we have
the analytic result that for $x < (E_{JT}/D_0)^2$ the effective
half bandwidth $D = \sqrt{x}D_0$, the localized $\ell$ levels lie
lower than the $b$ band bottom, only the former are occupied and
the system is an insulator.  The $T=0$ electrical gap  between the
occupied $\ell$ levels and the unoccupied $b$ band bottom in this
ferro-insulator phase  is given by $\Delta = E_{JT} -D = E_{JT}
-\sqrt{x} D_o$. The critical doping for the $T=0$ ferro-insulator
to ferro-metal transition, determined by the vanishing of
$\Delta$, is thus $x_c = (E_{JT}/D_0)^2$.  As $x$ increases beyond
$x_c$, the system becomes a ferromagnetic metal, with
$\epsilon^*_\ell$ the re-normalized $\ell$ level lying above the
$b$ band bottom. $\bar{n}_b$ increases with $x$ ,till at some
value $x_{c2} ,\,\bar{n}_b = (1-x_{c2})$ so that $\bar{n}_\ell
=0$. Beyond $x_{c2}$, only the $b$ band is occupied.

Thus our theory leads naturally to an insulating (ferromagnetic)
state for $x< x_c$, a ferro-metallic state with coexisting band
$b$ and localized $\ell$ electrons for $x_c < x < x_{c2}$, and a
metal with only $b$ electrons and bandwidth $2D_0$ for $x >
x_{c2}$. Indeed, for $x$ close to 1 ("electron doped limit") there
are surprisingly successful calculations \cite{pai-mag} of
(magnetic) ground states based on a model of independent broad
band tight binding electrons moving in appropriate magnetic
superstructures of $t_{2g}$ spins which have AF super-exchange
interactions, completely ignoring JT interactions. Our theory
provides a rationalization for this.

Detailed zero temperature results from our theory are shown in
Figs. 1(a)-(c) and in Fig. 2. We choose $E_{JT} = 0.5 \, eV$ and
$D_0 = 1 eV$. Fig. 1(a) shows the variation of the re-normalized
$\ell$ level and the $b$ band edge positions ($U=\infty$, full
line, $U= 5 eV$, a realistic value, dotted line) with doping $x$.
The effective $b$ bandwidth $2D$ becomes very small (0 for $U =
\infty$) as $x \rightarrow 0$. The physical reason is that, as
mentioned before, the $b$ electrons reside mostly on the hole
sites(fraction $x$), being strongly repelled from those occupied
by $\ell$ polarons  (with repulsion energy $U>>2D$). At $x = x_c$
($=0.25$ for $U=\infty$, and very nearly this value for $U= 5 \,
eV$), the band bottom crosses the $\ell$ level. The system is
metallic for $x > x_c$, with both $\bar{n}_b$ and $\bar{n}_\ell$
being nonzero, as shown in Fig. 1(b). We note that in the
`metallic' regime $x_c < x< 0.5,\, \bar{n}_b$, the average band
occupancy, is small, eg. at $x=0.4,\, \bar{n}_b \simeq \,.08$. For
the set of parameters chosen, the $\ell$ level empties out
completely for $x = x_{c2} \simeq 0.72$, and beyond this only band
states are occupied. Double exchange and anti-ferromagnetic
super-exchange describe the magnetic behaviour of the system in
this regime.  In Fig. 1(c), the effective band or electrical gap
is shown as a function of $x$. The gap vanishes smoothly as
$x\rightarrow x_c$ and rises as $x\rightarrow 0$ to a value
$E_{JT}$, and not $U$; this should be the electrical gap seen at
any finite $x$ no matter how small, or in a $T\neq 0$ experiment.
Fig. 2 shows how the $T=0$ insulator metal boundary ($x_c$) shifts
as a function of $D_0$ and $U$. $x_c$ increases as $D_0$
decreases, and as $U$ increases. We note that all physical
quantities for $U= 5eV$ are close to those for $U=\infty$.
\begin{figure}
\twoimages[width=6.5cm,height=9.0cm]{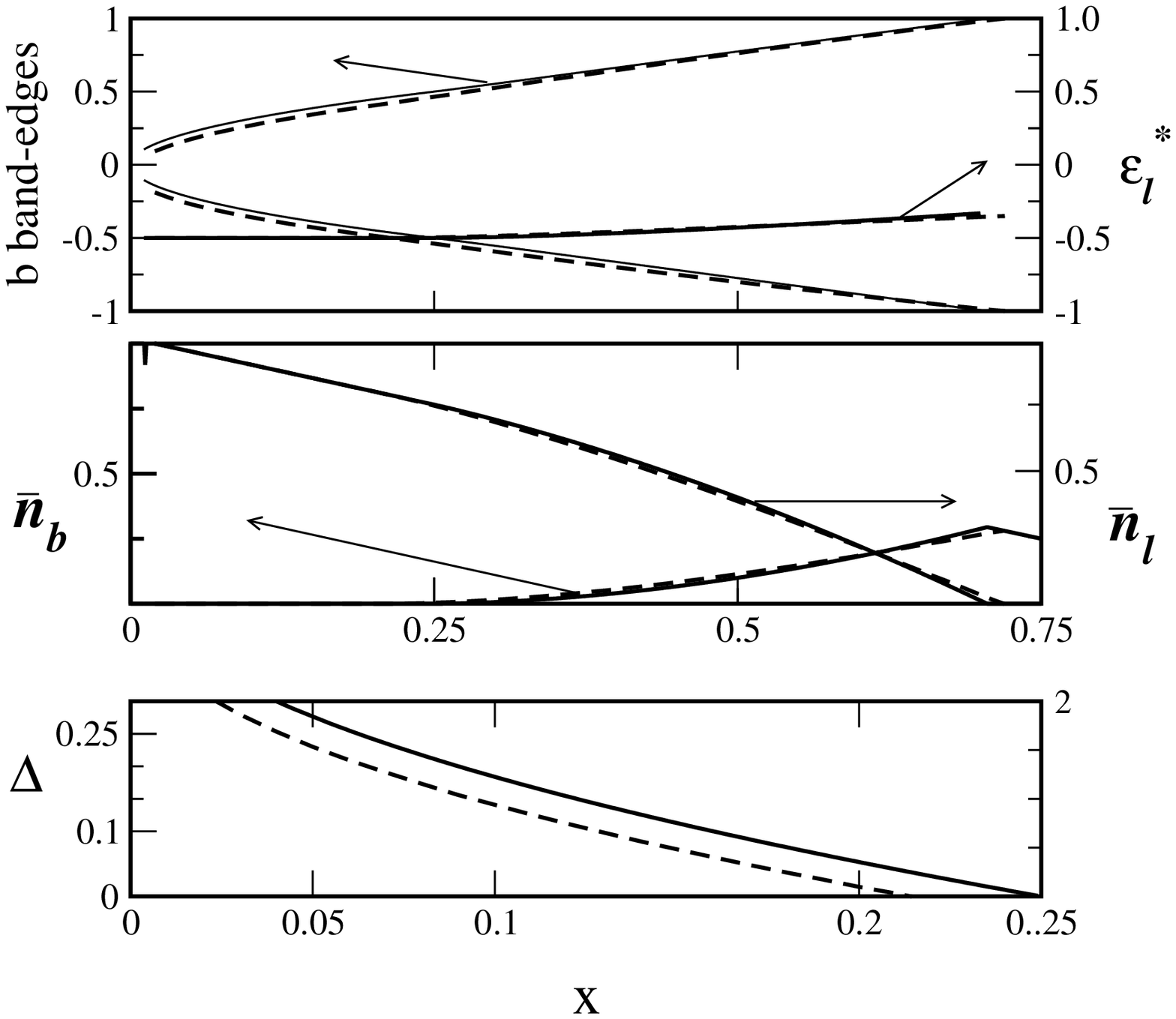}{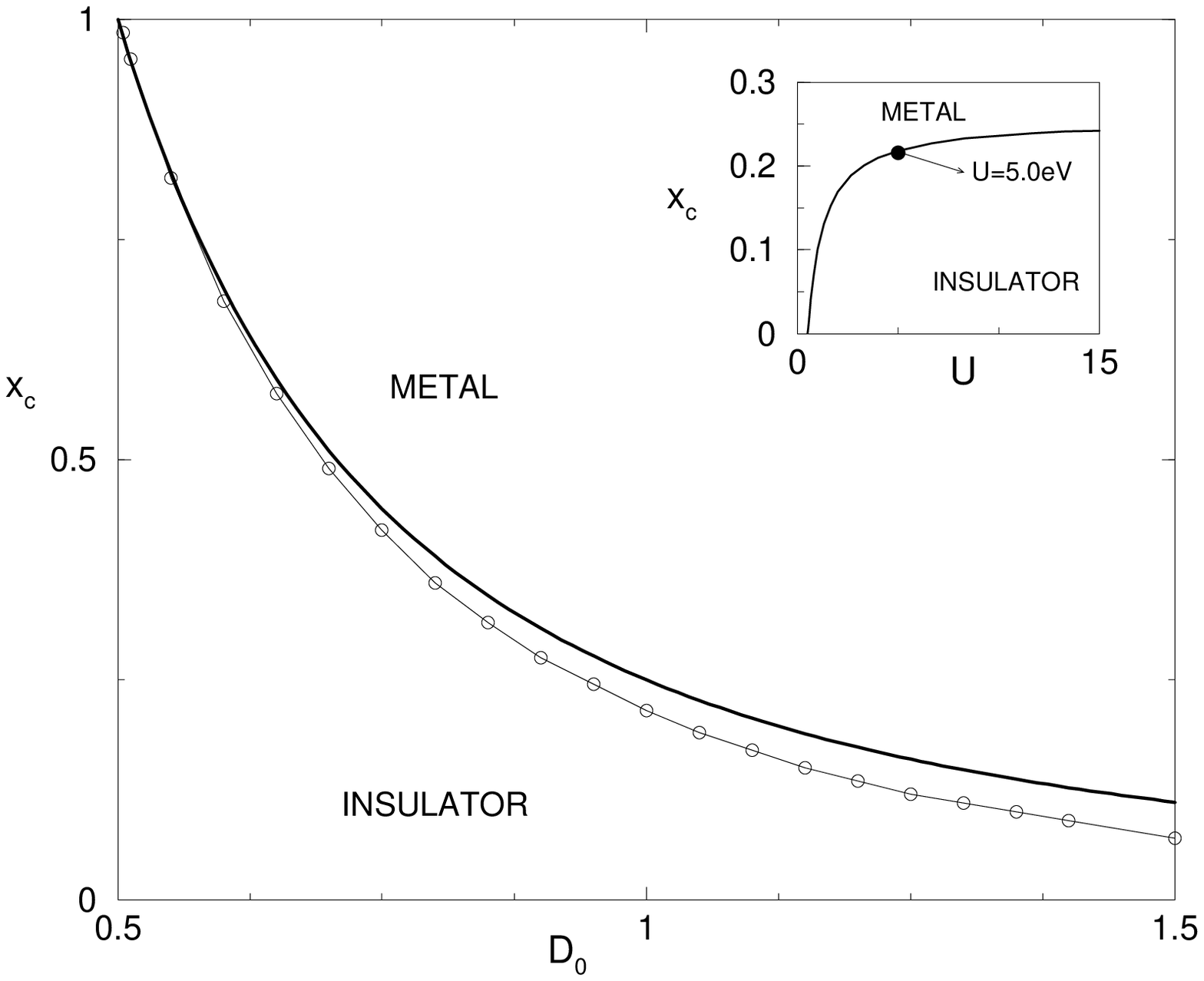}
\caption{(a) Variation of $b$ band edges and the effective $\ell$
level $\epsilon^*_\ell$, in units of $D_{0}$, with doping $x$. (b)
The number ${\bar n}_{l}$ (of localized JT polaronic $\ell$
electrons) and ${\bar n}_{b}$ ( of band electrons), per site, as a
function of $x$. (c) The $T=0$ insulating gap $\Delta \equiv
(E_{JT} - D)$, as a function of $x$. In all the cases $E_{JT}=0.5
\, eV$ and $D_{0} = 1 \, eV$. The full line corresponds to
$U=\infty$ and long dashed line to $U = 5 \, eV$.} \label{Fig1}
\vspace{-.3cm}

\caption{The critical concentration $x_c$ for the $T=0$ insulator
metal transition as a function of $D_{0}$ the bare $b$
half-bandwidth, in $eV$. The full line corresponds to  $U=\infty$
and line with circles to $U=5 \, eV$ . The inset shows $x_c$ as a
function of $U/D_{0}$ for $E_{JT}=0.5 \, eV$ and $D_{0}=1 \, eV$.
The point corresponding to $U=5 \, eV$ is marked by a filled
circle.} \label{Fig2} \vspace{-.3cm}
\end{figure}

We now briefly compare our results with observed material trends
and experimental numbers where available. One of our predictions
is that of a ferromagnetic but insulating ground state (FI) for $x
< x_{c}$. The former arises from the new virtual double-exchange
coupling $J_F$ described earlier; the state is insulating because
the effective $b$ half bandwidth $D < E_{JT}$ for $x < x_{c}$. In
contrast, the ferromagnetic state is necessarily metallic if it
arises solely due to double exchange\cite{d-e}. Experimentally,
all doped manganites have an insulating, fully polarized
ferromagnetic state, for $x_{c1}< x < x_{c}$ ; {\it eg.} for $La
Ca$, $x_{c1}=0.10$ and $x_{c}=0.18$. In our calculations,
$x_{c1}=0$ (because we have neglected the small, orbital order
dependent, anti-ferromagnetic exchange important at small $x$
\cite{comment1}), while the critical doping $x_c$ for the
insulator-metal transition at $T=0$ depends on material parameters
roughly as $x_c=(E_{JT}/D_{0})^2$ (cf. Fig. 2). This prediction
can not be directly compared with experiment since the systematics
of $E_{JT}$ and $D_{0}$ are not precisely known. It is however
believed \cite{radaelli} that the bare half-bandwidth $D_0$
decreases in the sequence $LaSr$, $LaCa$, $NdSr$ (and $PrCa$),
because of cation size and its effect primarily on the Mn-O-Mn
bond angle and via this on the nearest neighbour hopping, while
$E_{JT}$ does not change much. The observed $x_c$ for this
sequence has values $0.16$, $0.18$, and $\stackrel {>}{\sim}
0.20$, the trend being consistent with our prediction. The puzzle
as to why some manganites (eg. $PrCa$) have only insulating ground
states unlike the above can be understood within our theory in
terms of the characteristic values of $(E_{JT} /D_0)$ appropriate
to the materials (eg., from Fig. 2, for $E_{JT} \sim .5 \, eV$,
the ferro-insulator extends up to $x=.5$ if $D_0 \stackrel <{\sim}
0.7 \, eV$).

The electrical activation energy in the Ferro-insulator state goes
as $\Delta_{eff}=D_{0}(\sqrt{x_c}-\sqrt{x})$ for large $U$ in our
theory. This dependence cannot be compared with experiment, since
there are no measurements of activation energies close to $T=0$;
the only experimental results we are aware of \cite{mandal} are
for $T>T_{c}$, in the paramagnetic phase. The corresponding gaps
are not expected to go to zero at $x_c$. However, this high $T( >
T_c)$ activation energy does decrease as $x \rightarrow x_c$ as
expected from our theory.

An additional consequence of our model is that in the
ferromagnetic metallic ground state, the concentration of  mobile
($b$) electrons is rather small (Fig. 1(b)), and {\em {not}}
$(1-x)$, the total number of $e_g$ electrons per site. This is
exactly the inference from the small Drude weight, {\it i.e.} the
integrated optical conductivity, which is a direct measure of the
effective number of carriers. For example, Okimoto {\it et.
al.}\cite{Neff} find $n_{eff}\simeq 0.06$ for $LaSr$ at $x=0.3$.
Our results for  ${\bar n}_{b}$ quoted above are very much in this
range of smallness. A related consequence of our theory, arising
from the fact that the large majority of the $e_g$ electrons are
in polaronic $\ell$ states even in the metal, is the persistence
of local polaronic distortions well into the metallic phase, for
which there is considerable experimental
evidence\cite{polaron_expt}. Thus our results provide a natural
explanation for several unusual ground state properties of
manganites.

In summary, we have presented here a new coexisting polaron/broad
band electron model for manganites, which revives the
Falicov-Kimball model in a new, hitherto unexpected and
unexplored, setting. We have completely solved the model in this
new context within the DMFT, and shown that this leads to a
physical explanation and a quantitative theory of many
characteristic and hitherto puzzling ground state properties of
doped manganites in the doping regime $0.1<x< 0.5$.

SRH and GVP thank the Council of Scientific and Industrial
Research (India) for partial financial support. HRK would like to
acknowledge support from the Indo-French Centre for the Promotion
of Advanced Research through its grant no. 2400-1.


\begin{thebibliography}{99}
\bibitem{comment1}
We do not address the phase transition from the ferro-insulator to
the anti-ferro insulator as $x$ decreases, which requires a
careful treatment of competing exchange interactions and orbital
ordering effects present at small $x$, in this paper.
\bibitem{real-band}
A comparison with more detailed calculations with realistic
(Koster-Slater) anisotropic hopping, which we have carried out
\cite{tvr-nat,theses}, shows that the results do not change much
if $D_0$ above is matched with the half bandwidth of the realistic
band-structure.
\end{thebibliography}
\end{document}